\documentclass[12pt, draftclsnofoot, onecolumn]{IEEEtran}

\usepackage[T1]{fontenc}
\usepackage[english]{babel}
\usepackage[utf8]{inputenc}
\usepackage{amsmath}
\usepackage{amsfonts}
\usepackage{amssymb}
\usepackage{balance}
\usepackage{graphicx}
\usepackage{epstopdf}
\usepackage{epsfig}
\usepackage{tikz}
\usepackage{soul}
\usetikzlibrary{shapes,arrows}
\usepackage{multirow}

\usepackage{algorithm}
\usepackage{color}
\usepackage{pifont}
\usepackage{verbatim,mathrsfs, cite}
\usepackage[acronym,shortcuts]{glossaries}
\usepackage{setspace, url}
\usepackage{bm}

\makeglossaries
\newacronym{5G}{5G}{fifth generation}
\newacronym{AF}{AF}{amplify and forward}
\newacronym{AN}{AN}{artificial noise}
\newacronym{AWGN}{AWGN}{additive white Gaussian noise}
\newacronym{BS}{BS}{base station}
\newacronym{BPSK}{BPSK}{binary phase shift keying}
\newacronym{CAS}{CAS}{coding across subchannel}
\newacronym{ACBCA}{A-CBCA}{asymmetric channel-based cryptographic authentication}
\newacronym{CDF}{CDF}{cumulative distribution function}
\newacronym{CF}{CF}{compute and forward}
\newacronym{CP}{CP}{cyclic prefix}
\newacronym{CPS}{CPS}{coding per subchannel}
\newacronym{CSIT}{CSIT}{channel-state information at the transmitter}
\newacronym{DF}{DF}{decode and forward}
\newacronym{DFT}{DFT}{discrete Fourier transform}
\newacronym{FD}{FD}{full-duplex}
\newacronym{FBMC}{FMBC}{filterbank modulation}
\newacronym{GSVD}{GSVD}{generalized \ac{SVD}}
\newacronym{GFDM}{GFDM}{generalized frequency division multiplexing}
\newacronym{HD}{HD}{half-duplex}
\newacronym{IDA}{ID-A}{identification association}
\newacronym{IDV}{ID-V}{identification verification}
\newacronym{IDFT}{IDFT}{inverse discrete Fourier transform}
\newacronym{iid}{i.i.d.}{independent and identically distributed}
\newacronym{kkt}{KKT}{Karush-Kuhn-Tucker}
\newacronym{KL}{KL}{Kullback-Leibler}
\newacronym{IoT}{IoT}{Internet of things}
\newacronym{LLR}{LLR}{log likelihood ratio}
\newacronym{MAC}{MAC}{multiple access channel}
\newacronym{MI}{MI}{mutual information}
\newacronym{MIMO}{MIMO}{multiple-input multiple-output}
\newacronym{MIMOME}{MIMOME}{MIMO multiple-Eve}
\newacronym{ML}{ML}{maximum likelihood}
\newacronym{MMSE}{MMSE}{minium mean square error}
\newacronym{MRC}{MRC}{maximal ratio combining}
\newacronym{MSE}{MSE}{mean square error}
\newacronym{MT}{MT}{mobile terminal}
\newacronym{OFDM}{OFDM}{orthogonal frequency division multiplexing}
\newacronym{OFDMA}{OFDMA}{orthogonal frequency division multiple access}
\newacronym{OQAM}{OQAM}{offeset \ac{QAM}}
\newacronym{PC}{PC}{pilot contamination}
\newacronym{PDF}{PDF}{probability density function}
\newacronym{PLA}{PLA}{physical layer authentication}
\newacronym{PLS}{PLS}{physical layer security}
\newacronym{PSA}{PSA}{probability of successful attack}
\newacronym{QAM}{QAM}{quadrature amplitude modulation}
\newacronym{RB}{RB}{random binning}
\newacronym{SAR}{SAR}{secure authentication rate}
\newacronym{SCBCA}{S-CBCA}{symmetric channel-based cryptographic authentication}
\newacronym{SCF}{SCF}{scaled compute-and-forward}
\newacronym{SC-FDMA}{SC-FDMA}{single-carrier frequency division multiple access}
\newacronym{SKA}{SKA}{secret key agreement}
\newacronym{SKT}{SKT}{secret key throughput}
\newacronym{SWIPT}{SWIPT}{{\em secure} wireless information and power transfer}
\newacronym{LEP}{LEP}{linear Eve's processing}
\newacronym{SOP}{SOP}{secrecy outage probability}
\newacronym{SVD}{SVD}{singular value decomposition}
\newacronym{SNR}{SNR}{signal to noise ratio}
\newacronym{UA}{UA}{user authentication}
\newacronym{ZMUPCG}{ZMUPCG}{zero-mean unitary power complex Gaussian}

\newtheorem{theorem}{Theorem}
\newtheorem{lemma}{Lemma}

\newtheorem{remark}{Remark}
\newtheorem{definition}{Definition}

\title{Analysis of Channel-Based User Authentication \\ by Key-Less and Key-Based Approaches}

 \author{
	\IEEEauthorblockN{Stefano Tomasin \\ {\small Department of Information Engineering, University of Padova, Italy}}
}

\begin{document}
\maketitle

\begin{abstract}
User authentication (UA) supports the receiver in deciding whether a message comes
from the claimed transmitter or from an impersonating attacker. In cryptographic approaches 
messages are signed with either an asymmetric or symmetric key, and a source of randomness
is required to generate the key. In physical layer authentication (PLA) instead the receiver checks if
received messages presumably coming from the same source undergo the same channel. We compare these solutions by considering the physical-layer channel features as randomness
source for generating the key, thus allowing an immediate comparison with PLA (that already uses these
features). For the symmetric-key approach we use secret key agreement, while
for asymmetric-key the channel is used as entropy source at the transmitter. We focus
on the asymptotic case of an infinite number of independent and identically distributed
channel realizations, showing the correctness of all schemes and analyzing the secure authentication
rate, that dictates the rate at which the probability that UA security is broken goes to zero as
the number of used channel resources (to generate the key or for PLA) goes to infinity. Both passive
and active attacks are considered and by numerical results we compare the various systems.
\end{abstract}

\begin{IEEEkeywords}
Physical Layer Authentication; Physical Layer Security; Rayleigh Fading;  User Authentication.
\end{IEEEkeywords}

\glsresetall

\section{Introduction}

\Ac{UA} methods in communication systems are used to confirm the identity of a message sender \cite{Stallings}. In particular, the receiver must take a decision on who has transmitted the message considering that an attacker aims at impersonating the legitimate transmitter. Typically, \ac{UA} includes two phases: an {\em \ac{IDA}} phase, when the legitimate transmitter is assigned an identifying feature using an authenticated channel, and an {\em \ac{IDV}} phase, when the identifying feature is verified upon reception of a message. 

Many commonly-used \ac{UA} protocols for communication systems use as identifier a key, i.e., a secret know by either only the transmitter (asymmetric key) or both the transmitter and the receiver (symmetric key), and encryption techniques are adopted in the \ac{IDV} phase \cite{Stallings}. These methods go also under the name of key-based \acp{UA} \cite{Simmons-1985, Simmons-1988}. An alternative key-less approach is \ac{PLA} \cite{7270404}, in which the identifying feature is the physical channel over which the communication occurs. In this case the \ac{IDA} phase consists in the identification of the channel features, while the \ac{IDV} phase consists in checking if the received message has undergone the same channel of the \ac{IDA} phase (see e.g. \cite{6204019, 1607573,Xiao09} and \cite{7270404} for a survey). For example, in wireless systems, the propagation phenomena (e.g., fading) are associated to the specific position of both transmitter and receiver, thus the attacker should be in the same position of the legitimate transmitter for a successful impersonation. \ac{PLA} has been studied for discrete memory-less channels in \cite{6807688}; implementations for \ac{MIMO} and intersymbol-interference channels have been proposed in \cite{6204019} and \cite{7420748}, while a game-theoretic study of \ac{PLA} has  appeared in \cite{7815442}. Recently, a review of key-based and key-less security approaches with a comparison of cryptography and physical-layer security solutions has been proposed in \cite{7393435}. In \cite{6807688} the keyless \ac{UA} was studied in a new noisy model, where there are two  discrete memoryless channels, one  from sender to receiver and and the other the from adversary to the receiver, in addition to an insecure noiseless channel between the legitimate parties. Optimality of the scheme was further proofed in \cite{7047899}.

In both key-based and key-less approaches a source of randomness is needed to either generate the keys or identify the user feature. In this paper we focus on the use of physical-layer channel features as randomness source for all the \ac{UA} techniques. While the exploitation of channel features is intrinsic in \ac{PLA}, for key-based solutions we extract a random number (the key) from the channel. The key must remain secret to the attacker, in order to prevent impersonation: thus for symmetric-key systems we use the \ac{SKA} procedure of \cite{243431, 256484}, while for asymmetric-key systems the channel is used as an entropy source for one user only, and a set of secret bits is obtained from this source.

Most of the literature has considered separately key-based and key-less \ac{UA}. In \cite{Tomasin-wsa17} a first comparison of the approaches when the unique source of randomness was the channel has been proposed, but limited to quantized channel model with a finite number of samples per frame. In this paper instead we focus on the asymptotic case of an infinite number of available \ac{iid} channel realizations. In this asymptotic regime all considered \ac{UA} schemes can be shown to be correct, i.e., authentic  messages are always accepted. We then analyze the \ac{SAR}, that dictates the rate at which the probability that \ac{UA} security is broken goes to zero as the number of channel resources used to extract either the key or the channel features goes to infinity. In particular, we focus on time-invariant channels, time-variant channels and Rayleigh fading channels with \ac{AWGN}. Most derivations are obtained considering a passive attacker that listens to transmissions in the \ac{IDA} phase and then attempts to impersonate the legitimate transmitter in the \ac{IDV} phase. However, we also consider active attacks aiming at disrupting the \ac{IDA} phase in order to ease the impersonation attack in the \ac{IDV} phase. Beyond considering optimal attacks by Eve we also discuss a simple approach in which the attacker first performs a linear combination of all channel estimates to obtain the \ac{MMSE} linear estimate of the legitimate channel which is then used for the attack. By numerical results we closely compare the various systems, highlighting their potentials and vulnerabilities. 

The rest of the paper is organized as follows. Section II introduces the system model, with emphasis on both the channel and the considered schemes. The analysis of the protocols in terms of \ac{SAR} is performed in Section III, while its  derivation for the relevant case of reciprocal Rayleigh fading \ac{AWGN} channels is discussed in Section IV. Active attacks during the \ac{IDA} phase are discussed in Section V. Numerical results comparing the various strategies are presented in Section VI, before the main conclusions are outlined in Section VII.

{\em Notation}: $\mathbb{P}[{\mathcal X}]$ denotes the probability of the event $\mathcal X$. $\mathbb{E}[x]$ denotes the expectation of the random variable $x$, $\mathbb{H}(x)$ denotes the entropy of the discrete random variable $x$, and  $\bar{\mathbb{H}}(x)$ denotes the differential entropy of the continuous random variable $x$. $\mathbb{I}(x;y)$ denotes the mutual information between random variables $x$ and $y$. ${\mathbb D}(p||q)$ denotes the \ac{KL} divergence between the two \ac{PDF} $p$ and $q$. Vectors are denoted in boldface, while scalar quantities are denoted in italic. $\ln x$ and $\log x$ denote the natural-base and base-2 logarithms of $x$, respectively. $\det \bm{A}$ and ${\rm tr} \bm{A}$ denote the determinant and trace of matrix $\bm{A}$, respectively. $p_x(a)$ denotes the \ac{PDF} of random variable $x$.

\section{System Model}
\label{chmodel}

We consider a communication system with two legitimate users, Alice and Bob, and one attacker Eve. Time is divided into frames, each of the same duration. In order to simplify the analysis we suppose that in even and odd frames Alice and Bob alternate their transmissions. In particular, starting from the first frame and in all odd frames $2t+1$ ($t=0, \ldots$) Alice transmits to Bob, while starting from the second frame and in all even frames $2t$ ($t=1, \ldots$) Bob transmits to Alice. Eve can transmit at any frame.

Bob must decide if packets received in odd frames are coming from either Alice or Eve. To this end, initial frames are used for \ac{IDA} purposes, extracting keys from the channel or identifying reference channel features as described in the following. In forthcoming frames Bob receives packets and performs \ac{IDV}, i.e., he decides if they are coming from Alice (hypothesis $\mathcal H_0$) or not (hypothesis $\mathcal H_1$). 

Channel between each users' couple are assumed to be time-invariant within each frame, and described by $n$ complex random variables. All transmissions include pilot symbols known to all users (including Eve) for channel estimation.  The channel estimated by Bob at frame $2t+1$ is described by the random vector 
\begin{equation}
\bm{X}(2t+1) = [x_1(2t+1), \ldots, x_n(2t+1)],
\end{equation}
while the channel estimated by Alice in frame $2t$ is described by the random vector
\begin{equation}
\bm{Y}(2t) = [y_1(2t), \ldots, y_n(2t)].
\end{equation}
Let 
\begin{equation}
\bm{V}_{\rm A}(2t+1) = [v_{1, {\rm A}}(2t+1), \ldots, v_{n, {\rm A}}(2t+1)]
\end{equation}
\begin{equation}
\bm{V}_{\rm B}(2t) = [v_{1, {\rm B}}(2t), \ldots, v_{n, {\rm B}}(2t)]
\end{equation}
be the estimated channels by Eve when either Alice or Bob are transmitting, respectively. We also collect into the vector $\bm{Z}(2t)$ the channel estimates of Eve up to frame $2t$, i.e.,
\begin{equation}
\bm{Z}(2t) = [\bm{V}_{\rm A}(1), \bm{V}_{\rm B}(2), \ldots, \bm{V}_{\rm A}(2t-1), \bm{V}_{\rm B}(2t)]\,. 
\end{equation}
Eve will exploit all these estimates to perform her attacks. 

Channel estimates $\bm{X}(t_1)$, $\bm{Y}(t_2)$, $\bm{V}_{\rm A}(t_3)$ and $\bm{V}_{\rm B}(t_4)$ are assumed to be correlated, in general however they do not have the same value since they are affected by noise, time-varying channel phenomena and changing positions of transmitters and receivers. Within each estimated vector, we assume that $x_k(t)$, are \ac{iid} for $k \in [1, n]$, with frame-dependent \ac{PDF} $p_{x(t)}(a)$, $a \in \mathbb C$. Similarly we have that $y_k(t)$, $v_{k, \rm A}(t)$ and $v_{k, \rm B}(t)$ are \ac{iid}. In the following we will drop index $k$ from all channel estimates, i.e., using $x(t)$ instead of $x_k(t)$ (for any $k$). Moreover, we define the $2t$-size vector 
\begin{equation}
\bm{z}(2t) = [v_{\rm A}(1), v_{\rm B}(2), \ldots, v_{\rm A}(2t-1), v_{\rm B}(2t)]\,.
\end{equation}

We consider three \ac{UA} approaches, namely \ac{ACBCA}, \ac{SCBCA} and \acf{PLA}. While both \ac{ACBCA} and \ac{SCBCA} have a cryptography background and exploit some secret known by either or both the legitimate parties, \ac{PLA} has information-theoretic foundations and mostly relies on the communication channel characteristics. In order to obtain a fair comparison of the three approaches, we extract the secret needed for both \ac{ACBCA} and \ac{SCBCA} from the channel, so that all three schemes have the same source of randomness.

Moreover, both Alice and Bob must be able to exchange authenticated messages in the \ac{IDA} phase, to ensure that they are talking to each other, and not with Eve. This is a common feature to all \ac{UA} protocols and we assume it for granted. The authenticated channel can be obtained for example when devices are in a protected location immune from attacks. Also, the authenticated channel is available when the keys used for authentication must be renewed, but we can still use the old keys in the \ac{IDA} phase. Moreover \ac{PLA}, which is not based on a key, can be considered as a low-complexity alternative to be used in replacement of a more expensive \ac{UA} procedure that is adopted only for the \ac{IDA} phase of the \ac{PLA} protocol.

We now details the \ac{UA} procedures for the three authentication approaches.

\subsection{\Ac{ACBCA}} 

The basic structure of the \ac{ACBCA} can be summarized as follows.

\paragraph*{\ac{IDA} Phase} In this phase Alice generates a private and public key couple from a source of randomness, where the public key decrypts messages encrypted with the private key. Then, she transmits the public key to Bob over an authenticated channel.

\paragraph*{\ac{IDV} Phase} In this phase, whenever Alice wants to transmit a message, she encrypts it with her private key, including a signature. Bob decrypts the message with the public key and checks if the signature is correct, and in this case the message is accepted as authentic, otherwise it is discarded as non-authentic.

Suitable variants of this scheme have been proposed to prevent various attacks (including the replay attack) and are out the scope of this paper. With proper adaptations, the scheme is proofed to be secure \cite{Stallings}, provided that the private key  is secret. 

In our implementation of \ac{ACBCA}, the random bits are obtained from the channel. In particular, in the second frame Alice extracts random bits from $\bm{Y}(2)$. On her side, Eve attempts to extract the private key from her channel observations. Therefore, we must ensure that the randomness extraction procedure provides bit that are not known to Eve. Proof of correctness and security in this case is provided in Section \ref{proanalysis}.

\subsection{\Ac{SCBCA}} 

The basic structure of the \ac{SCBCA} can be summarized as follows.

\paragraph*{{\ac{IDA} Phase}} Alice and Bob share over an authenticated and confidential (to Eve) channel a secret key. 

\paragraph*{\ac{IDV} Phase} Alice encrypts the message with her secret key, including a signature. Bob decrypts the message with the same secret key and checks if the signature is correct. 

Also this scheme, with proper modifications and improvements (out of the scope of this work) is proofed to be secure \cite{Stallings}, provided that the key is secret. In our implementation of \ac{SCBCA} the key is obtained from the channel, using what is known as a \ac{SKA} protocol \cite{Bloch} that provides a secret key to two parties, with the support of an authenticated channel. In particular, in the first two frames they estimate their channels and using a public error-free authenticated channel they complete the reconciliation process. The proof of correctness and security in this case is provided in Section \ref{proanalysis}.

\subsection{\ac{PLA}}

With \ac{PLA} the authentication is provided by the physical channel over which the communication occurs \cite{6204019}. The \ac{PLA} algorithm works as follows.

\paragraph*{{\ac{IDA} Phase}} In this phase, occurring at frame 1,  Bob obtains a reference estimate of the channel to Alice $\bm{X}(1)$. The estimation must be performed using an authenticated message, so that Bob is sure that the estimated channel is connecting him to Alice (rather than Eve).

\paragraph*{{\ac{IDV} Phase}} In this phase, occurring for frames $2t+1$, with $t>0$, whenever Bob receives a message, he decides that it comes from Alice if the estimated channel in that frame, $\bm{X}(2t+1)$, is similar to $\bm{X}(1)$. 

The correctness and security proof of \ac{PLA} is provided in Section \ref{proanalysis}.

\section{Protocols Analysis With \ac{IDV} Attacks}
\label{proanalysis}

In this section we consider only the attacks in \ac{IDV} phase, while  attacks in the \ac{IDA} phase will be considered in Section \ref{attackana}. We assess the correctness and security (as usually defined in cryptography) of each \ac{UA} protocol. We first provide a definition of correctness and security:
\begin{definition}[Correctness and Security]
A \ac{UA} protocol is correct when a message coming from Alice is verified as authentic by Bob and is secure when a message coming from Eve is dismissed as non-authentic by Bob. 
\end{definition}

Let $\bm{K}(n)$ be the key used for key-based \ac{UA} or the features used for \ac{PLA}: since we assume that the $n$ channel realizations are \ac{iid}, the length of $\bm{K}(n)$ grows linearly with $n$. We aim at establishing the protocols' correctness and security asymptotically, as the number of channel uses per frame $n$ goes to infinity. About security, we show that for all schemes, the \ac{PSA} $P_s(t)$ at frame $t$ goes exponentially to zero as $n$ goes to infinity and we define
\begin{equation}
R(t) = \lim_{n\rightarrow \infty} - \frac{1}{n} \log P_s(t)
\label{Psdef}
\end{equation}
as the {\em \acf{SAR}} at frame $t$. Interpreting the $n$ entries of the channel estimate vectors as $n$ {\em channel uses}, $R(t)$ is the number of {\em secret bits in $\bm{K}(t)$ per channel use} at frame $t$, i.e, the number of bits that are not known to Eve in $\bm{K}(n)$, as $n\rightarrow \infty$. Note that the \ac{PSA} depends on the frame index $t$ since Eve collects channel estimates as time goes on, thus being able to refine her knowledge of the Alice-Bob channel. In the following we derive the maximum \ac{SAR} for each \ac{UA} protocol achieved by proper use of channel knowledge. We consider attacks at odd frames, starting from frame $t=3$, after Eve has collected an even number of channel observations. 

\subsection{\ac{SCBCA}} 

We establish the correctness and security of the \ac{SCBCA} protocol by the following Theorem.
\begin{theorem}
As $n\rightarrow \infty$, the \ac{SCBCA} protocol is correct and the \ac{SAR} for frame $2t+1$, with $t \geq 1$ is 
\begin{equation}
R_{\rm S-CBCA}(2t+1) = C_{\rm SKA}(2t)\,,
\label{rSCBCA}
\end{equation}
where $C_{\ac{SKA}}(2t)$ is the {\em weak secret-key capacity} of the \ac{SKA} process between Alice and Bob, given the channel knowledge by Eve at frame $2t$.   
\end{theorem}

\begin{IEEEproof}
\ac{SCBCA} is proofed to be correct and secure \cite{Stallings} provided that the key is secret.  About correctness,  asymptotically \ac{SKA} ensures that Alice and Bob have the same key, as long as the secret key rate $R_{\rm SKA}(2t+1)$ satisfies $R_{\rm SKA}(2t+1)  \leq C_{\rm SKA}(2t+1)$. 

About security, for \ac{SKA} the length of the secret key (in bits) grows as $n R_{\rm SKA}(2t+1)$. Therefore 
\begin{equation}
P_s(t) \approx 2^{-nR_{\rm SKA}(2t+1)}
\end{equation}
and from (\ref{Psdef}) the maximum \ac{SAR} coincides with $C_{\rm SKA}(2t+1)$, and we obtain (\ref{rSCBCA}). 
\end{IEEEproof}

We recall that the secret key capacity for the source-model \ac{SKA} is not known with a closed-form expression, but is bounded as
\begin{equation}
\begin{split}
{\mathbb I}(x(1);y(2)) - \min \{{\mathbb I}(x(1);\bm{z}(2t)), {\mathbb I}(y(2);\bm{z}(2t))  \} \\
\leq  C_{\ac{SKA}}(2t+1) \leq  \\ 
\min\{{\mathbb I}(x(1);y(2)), {\mathbb I}(x(1);y(2)|\bm{z}(2t)) \}\,,
\end{split}
\label{boundSKA}
\end{equation}
since the two channel estimates used for \ac{SKA} are $\bm{X}(1)$ and $\bm{Y}(2)$ and the information on the channel available at Eve at frame $2t+1$ is $\bm{Z}(2t)$.

\subsection{\ac{ACBCA}}

For \ac{ACBCA} we first consider the case in which the channel estimates are discrete-valued random variable. This occurs for example if Alice quantizes $y(2)$ into $\langle y(2)\rangle \in \mathcal A$, where $\mathcal A = \{a_0, \ldots, a_{M-1}\}$ is an $M$-size quantization alphabet. The random number used for the private key is then extracted from $\langle y(2)\rangle$. 

\begin{theorem}
The \ac{ACBCA} scheme is correct and for discrete-valued estimated channels $\langle y(2)\rangle$, as $n\rightarrow \infty$ its \ac{SAR} at frame $2t+1$, with $t\geq 1$ is 
\begin{equation}
R_{\rm A-CBCA}(2t+1) = \mathbb{H}(\langle y(2)\rangle|\bm{z}(2t))\,, 
\label{rACBCA}
\end{equation}
where $\mathbb{H}(\cdot|\cdot)$ is the conditional entropy. 
\end{theorem}

\begin{IEEEproof}
Correctness of \ac{ACBCA} scheme follows immediately by the assumption that the public-key broadcast is error-free and authenticated. 

About security, the \ac{ACBCA} protocol can be seen as a source-based \ac{SKA} protocol where both Alice and Bob observe the same source of randomness. In this case the \ac{SKA} capacity upper bound becomes
\begin{equation}
\begin{split}
\min\{{\mathbb I}(\langle y(2)\rangle;\langle y(2)\rangle), {\mathbb I}(\langle y(2)\rangle;\langle y(2)\rangle|\bm{z}(2t)) \} \\
= {\mathbb I}(\langle y(2)\rangle;\langle y(2)\rangle|\bm{z}(2t)) = \mathbb{H}(\langle y(2)\rangle|\bm{z}(2t))\,,
\end{split}
\end{equation}
which coincides with the lower bound 
\begin{equation}
\begin{split}
{\mathbb I}(\langle y(2)\rangle;\langle y(2)\rangle) -  {\mathbb I}(\langle y(2)\rangle;\bm{z}(2t))    \\
= \mathbb{H}(\langle y(2)\rangle) - \mathbb{H}(\langle y(2)\rangle) + \mathbb{H}(\langle y(2)\rangle|\bm{z}(2t))\,,
\end{split}
\end{equation}
providing (\ref{rACBCA}).
\end{IEEEproof}

For the channel quantization case a lower bound on \ac{SAR} is obtained as follows
\begin{equation}
\begin{split}
R_{\rm A-CBCA}(2t+1) & = \mathbb{H}(\langle y(2)\rangle) -  \mathbb I(\langle y(2)\rangle;\bm{z}(2t)) \\ 
& \geq \max\{0, \mathbb{H}(\langle y(2)\rangle) -  \mathbb I(y(2);\bm{z}(2t))\}\,,
\end{split}
\label{Hgaussbound}
\end{equation}
where the first line is obtained by the definition of mutual information and the second line by the data processing inequality, where we upper-bounded the mutual information on the quantized variable by the corresponding mutual information on the continuous-valued channel estimate. 

Now, we consider the case in which the channel is a continuous random random variable, for which the \ac{SAR} is established by the following Lemma.

\begin{lemma}
For continuous channel estimates at Alice (i.e., $y(2)$ is a continuous random variable), the \ac{SAR} is
\begin{equation}
R_{\rm A-CBCA}(2t+1) =\infty\,.
\label{pippo}
\end{equation}
\end{lemma}

\begin{IEEEproof}
We first observe that the conditional entropy in (\ref{rACBCA}) can be written as
\begin{equation}
\mathbb{H}(\langle y(2)\rangle|\bm{z}(2t)) = \int \mathbb{H}(\langle y(2)\rangle|\bm{z}(2t) = \bm{b}) p_{\bm{z}(2t)}(\bm{b}) {\rm d}\bm{b}\,. 
\end{equation}
Now, the continuous random variable $y(2)$ is the asymptotic case of the quantized $\langle y(2)\rangle$ with a number of quantization points $M \rightarrow \infty$. Asymptotically, the conditional entropy for each value of $\bm{z}(t)$ is the {\em limiting density of discrete points} which tends to infinity logarithmically with $M$ under very mild assumptions \cite{Jaynes-63}. Therefore, the average of the logarithm, which corresponds to the conditional entropy tends to infinity, i.e., $\mathbb{H}(\langle y(2)\rangle|\bm{z}(2t)) \rightarrow \infty$, providing (\ref{pippo}).
\end{IEEEproof}

\subsection{\ac{PLA}} 
\label{pla}

We establish the correctness and security of \ac{PLA} by the following Theorem.
\begin{theorem}
The \ac{PLA} protocol is asymptotically correct, and as $n\rightarrow \infty$ its \ac{SAR} at frame $2t+1$, with $t \geq 1$, is 
\begin{equation}
R_{\rm \ac{PLA}}(2t+1) = {\mathbb D}(p_{x(1),x(2t+1)|\mathcal H_0}||p_{x(1),x(2t+1)|\mathcal H_1}),
\label{rPLA1}
\end{equation}
where $p_{x(1),x(2t+1)|\mathcal H_0}$ and $p_{x(1),x(2t+1)|\mathcal H_1}$ are the joint \ac{PDF} of $x(1)$ and $x(2t+1)$ when Alice and Eve are transmitting, respectively in the \ac{IDV} phase (frame $2t+1$).
\end{theorem}

\begin{IEEEproof}
About correctness, using the Chernoff-Stein Lemma \cite[Theorem 11.8.3]{Cover} we have that for $n\rightarrow \infty$ the probability of correctly authenticating messages coming from Alice can be made arbitrarily small.


About security, still from the Chernoff-Stein Lemma \cite[Theorem 11.8.3]{Cover} the \ac{PSA} goes to zero as 
\begin{equation}
P_s \sim 2^{-n{\mathbb D}(p_{x(1),x(2t+1)|\mathcal H_0}||p_{x(1),x(2t+1)|\mathcal H_1})}\,,
\end{equation}
which directly provides (\ref{rPLA1}).
\end{IEEEproof}

Note that the \ac{SAR} depends on the attack that Eve is performing through $p_{x(1),x(2t+1)|\mathcal H_1}$. Therefore it is interesting to have results under the hypothesis that Eve performs a specific attack, as in the following Lemma.
\begin{lemma}
\label{lpr}
If Eve is able to induce any channel estimate $\bm{X}(2t+1)$ to Bob when attacking, and assuming that Eve generates the induced estimate randomly distributed according to the \ac{PDF} of the legitimate channel given Eve's observations, $p_{x(2t+1)|\bm{z}(2t), \mathcal H_0}$, then the \ac{SAR} is  upper-bounded by
\begin{equation}
R_{\rm \ac{PLA}}(2t+1) \leq {\mathbb I}(x(1);x_{\mathcal H_0}(2t+1)|\bm{z}(2t))\,.
\label{rPLA}
\end{equation}
\end{lemma}
\begin{IEEEproof}
See Appendix \ref{proof2}.
\end{IEEEproof}

From this lemma we observe that if the statistics of estimated channels at both Alice and Bob are the same ($p_x = p_y$) then $R_{\rm \ac{PLA}}(2t+1) \leq {\mathbb I}(x(1);y(2)|\bm{z}(2t))$, i.e., the \ac{PLA} has the same upper-bound of \ac{SCBCA}.

\begin{remark}
For time-varying channels we assumed that Eve can induce any channel estimate when performing the attack. This however does not hold as the channel are changing and Eve does not know the exact channel (with respect to Bob) on which she attacks, therefore in practice she will not be able to induce any desired channel estimate. Therefore our assumption is conservative and the obtained \ac{SAR} is a lower bound on the effective \ac{SAR}. Moreover, for \ac{PLA} the time-varying channel scenario is particularly challenging, since the variations decrease the ability of the receiver to identify the channel in the \ac{IDV} phase. In other words, for \ac{PLA} \ac{SAR} will decrease over time. In particular, as $t\rightarrow \infty$ the \ac{KL} divergence will tend to zero, thus nulling the \ac{SAR}. 
\end{remark}
\begin{remark}
We have assessed the performance for the three \ac{UA} methods, and from the theorems we can conclude that \ac{ACBCA} has  an unlimited \ac{SAR} as we increase the number of quantization bits. For \ac{PLA} we have an explicit expression for \ac{SAR}, depending however on the attack by Eve. For the \ac{SAR} of \ac{SCBCA} we have only bounds. Moreover, for a particular attack strategy by Eve the \ac{SAR} of \ac{PLA} share the same upper bound of \ac{SCBCA}.

\end{remark}
\section{Rayleigh AWGN Reciprocal Channels}
\label{secAWGN}

We now consider a scenario in which the estimates are corrupted by \ac{AWGN}, which corresponds for example to a massive \ac{MIMO} system in which all users (Alice, Bob and Eve) have $\sqrt{n}$ antennas each, so that the resulting channel matrices have $n \rightarrow \infty$ entries\footnote{Other antennas configurations can be considered, leading to similar results and expressions as those derived in this section. Also, \ac{OFDM} systems can be cast in this model.}. Moreover, channel matrix entries are assumed \ac{iid}, and \ac{ZMUPCG}, in accordance with the Rayleigh fading model. By reordering the $n$ entries of each matrix into a vector we obtain the channel model described in Section \ref{chmodel}, where $x(t)$, $y(t)$ and $z(t)$ are Gaussian distributed.

We assume that the channel between any couple of devices is reciprocal, therefore we have
\begin{equation}
x(t) = h(t) + \sigma_{\rm x} w_{\rm x}(t)\,, \quad y(t) = h(t) + \sigma_{\rm y} w_{\rm y}(t)\,, 
\label{xeq}
\end{equation}
where $w_{\rm x}(t), w_{\rm y}(t)$ are jointly \ac{ZMUPCG} and $\sigma_{\rm x}^2, \sigma_{\rm y}^2$ are the noise powers at the receivers. We also define the correlation of $h(t)$ over time as
\begin{equation}
\mathbb E[h(t)h^*(t+\ell)] = \rho(\ell)\,.
\end{equation}

Eve's channels are correlated with coefficients $\alpha_{\rm A} \in [-1, 1]$ and $\alpha_{\rm B} \in [-1, 1]$ to $h(t)$, and affected by \ac{AWGN} with powers $\sigma_{\rm v,A}^2$ and $\sigma_{\rm v,B}^2$ (typically $\sigma_{\rm v,A}^2  =  \sigma_{\rm v,B}^2$). Therefore, her estimate of the channel to Alice is 
\begin{equation}
v_{\rm A}(t)= \alpha_{\rm A} h(t) + \sqrt{1-\alpha_{\rm A}^2} q_{\rm A}(t) + \sigma_{\rm v, A}  w_{\rm v, A}(t)\,, 
\label{veqA}
\end{equation}
while her estimate of the channel to Bob is
\begin{equation}
v_{\rm B}(t) = \alpha_{\rm B} h(t) + \sqrt{1-\alpha_{\rm B}^2} q_{\rm B}(t) + \sigma_{\rm v, B}  w_{\rm v, B}(t)\,,
\label{veqB}
\end{equation}
where $w_{\rm v, A}(t)$, $w_{\rm v, B}(t)$  are \ac{ZMUPCG} independent with respect to $h(t)$, $q_{\rm A}(t)$ and $q_{\rm B}(t)$, and $q_{\rm A}(t)$, $q_{\rm B}(t)$ are \ac{ZMUPCG} variables. Note that by these definitions all channel estimates have unitary variance in the absence of noise.

\paragraph*{\ac{ACBCA}}  As we have already seen, using directly the continuous-valued channel estimate we obtain that the \ac{SAR} of \ac{ACBCA} is infinite, thus we focus here on the channel quantization case. The conditional entropy in (\ref{rACBCA}) can be written (by definition) as 
\begin{equation}
\begin{split}
R&_{\rm A-CBCA}(2t+1) = \mathbb{H}  (\langle y(2)\rangle|\bm{z}(2t)) = \mathbb{H}  (\langle y(2)\rangle|\bm{z}(2t)) =  \\
&-\sum_{i=0}^{M-1} \int p_{\langle y(2)\rangle|\bm{z}(2t)}(a_i|\bm{b}) p_{\bm{z}(2t)}(b) \log p_{\langle y(2)\rangle|\bm{z}(2t)}(a_i|\bm{b})  {\rm d}\bm{b}\,.
\end{split}
\end{equation} 
About the bound (\ref{Hgaussbound}),  recall that for a Gaussian vector $\bm{v}$ of size $k$ with correlation matrix $\bm{R}_{\rm v}$ the differential entropy is $\bar{\mathbb H}(\bm{v}) = \log \det ((\pi e)^k \bm{R}_{\rm v})$. Let us define
\begin{equation}
\begin{split}
\bm{R}&_{[y(2), \bm{z}(2t)]} = \\
=& \mathbb E[[y(2), \bm{z}(2t)]^H[y(2), \bm{z}(2t)]]  \\ 
= &  \left[\begin{matrix}
1 +\sigma_{\rm y}^2& \bm{r}_{\rm y}^H \\
\bm{r}_{\rm y} & \bm{R}_{\bm{z}(2t)}
\end{matrix}\right]\,,
\end{split}
\end{equation}
where $\bm{r}_{\rm y} = \mathbb E[y(2)\bm{z}^H(2t)]$ has entries
\begin{equation}
[\bm{r}_{\rm y}]_\ell = \begin{cases}
\alpha_{\rm A}^*\rho(\ell - 2) & \ell \mbox{ even}\,, \\
\alpha_{\rm B}^*\rho(\ell - 2) & \ell \mbox{ odd}\,. \\
\end{cases}
\end{equation}
Then we have 
\begin{equation}
\begin{split}
\mathbb I& (y(2);\bm{z}(2t)) =  \bar{\mathbb H}(y(2)) + \bar{\mathbb H}(\bm{z}(2t)) \\ & - \bar{\mathbb H}([y(2), \bm{z}(2t)]) 
 =  \log \frac{(1 + \sigma_{\rm y}^2)\det \bm{R}_{\bm{z}(2t)}}{\det \bm{R}_{[y(2), \bm{z}(2t)]}} \\
 & =  - \log \left(1 - \frac{\bm{r}_{\rm y}^H\bm{R}_{\bm{z}(2t)}^{-1} \bm{r}_{\rm y}}{1 + \sigma_{\rm y}^2} \right) \,,
\end{split}
\label{Iyz}
\end{equation}
where $\bm{R}_{\bm{z}(2t)} = \mathbb E[\bm{z}^H(2t)\bm{z}(2t)]$, with entries
\begin{equation}
[\bm{R}_{\bm{z}(2t)}]_{m,n} = \begin{cases}
\alpha_{\rm A}^*\alpha_{\rm B}\rho(m-n) & \mbox{$n$ odd, $m$ even} \\
\alpha_{\rm A}\alpha_{\rm B}^*\rho(m-n) & \mbox{$n$ even, $m$ odd} \\
|\alpha_{\rm B}|^2\rho(m-n) & \mbox{$n$ and $m$ even, $m\neq n$} \\
|\alpha_{\rm A}|^2 \rho(m-n) & \mbox{$n$ and $m$ odd, $m\neq n$} \\
1 +\sigma_{\rm z}^2 & n=m\,.
\end{cases}
\end{equation}

\paragraph*{\ac{SCBCA}} For \ac{SCBCA} we have the bound (\ref{boundSKA}). In particular for the Gaussian case we have 
\begin{equation}
\begin{split}
\mathbb I(x(1);y(2)) & = \mathbb H(x(1)) + \mathbb H(y(2)) - \mathbb H(x(1), y(2))  \\
& = -\log\left(1 - \frac{|\mathbb E[x(1)y^*(2)]|^2}{(1 + \sigma_{\rm x}^2)(1+ \sigma_{\rm y}^2)} \right) \\
& = -\log\left(1 - \frac{\rho(1)}{(1 + \sigma_{\rm x}^2)(1+ \sigma_{\rm y}^2)} \right)\,,
\end{split}
\end{equation}
and analogously to (\ref{Iyz})
\begin{equation}
\begin{split}
{\mathbb I}(x(1);\bm{z}(2t)) = & \log \frac{(1 + \sigma_{\rm x}^2)\det \bm{R}_{\bm{z}(2t)}}{\det \bm{R}_{[x(1), \bm{z}(2t)]}} \\
& = - \log \left(1 - \frac{\bm{r}_{\rm x}^H\bm{R}_{\bm{z}(2t)}^{-1} \bm{r}_{\rm x}}{1 + \sigma_{\rm x}^2} \right) 
\end{split}
\end{equation}
where 
\begin{equation}
\begin{split}
\bm{R}&_{[x(1), \bm{z}(2t)]} \\
=& \mathbb E[[x(1), \bm{z}(2t)]^H[y(2), \bm{z}(2t)]]  \\ 
= &  \left[\begin{matrix}
1 +\sigma_{\rm x}^2 & \bm{r}_{\rm x}^H\\
\bm{r}_{\rm x} & \bm{R}_{\bm{z}(2t)}
\end{matrix}\right]\,,
\end{split}
\end{equation}
\begin{equation}
[\bm{r}_{\rm x}]_\ell = [\mathbb{E}[x(1)\bm{z}^H(2t)]]_\ell = \begin{cases}
\alpha_{\rm A}^*\rho(\ell - 1) & \ell \mbox{ even}\,, \\
\alpha_{\rm B}^*\rho(\ell - 1) & \ell \mbox{ odd}\,. \\
\end{cases}
\end{equation}
Moreover we have
\begin{equation}
{\mathbb I}(x(1);y(2)|\bm{z}(2t)) = {\mathbb I}(x(1);y(2),\bm{z}(2t)) - {\mathbb I}(x(1);\bm{z}(2t))
\end{equation}
\begin{equation}
{\mathbb I}(x(1);y(2),\bm{z}(2t)) = \log\frac{(1+\sigma_{\rm x}^2)  \det \bm{R}_{[y(2),\bm{z}(2t)]}}{\det \bm{R}_{[x(1), y(2),\bm{z}(2t)]}} 
\label{Ixzdz}
\end{equation}
and
\begin{equation}
\begin{split}
 \bm{R}&_{[x(1), y(2),\bm{z}(2t)]} = \\
& \left[\begin{matrix}
 1+\sigma_{\rm x}^2 & \rho(1) & \bm{r}_{\rm x}^H \\
\rho(-1) & \multicolumn{2}{c}{\multirow{2}{*}{$\bm{R}_{[y(2),\bm{z}(2t)]}$}} \\
 \bm{r}_{\rm x} &  & \\
 \end{matrix}\right]\,.
	\end{split}
\end{equation}

\paragraph*{\ac{PLA}} Assuming that Eve generates the induced estimate $x(2t+1)$ randomly distributed according to the \ac{PDF} of the legitimate channel given Eve's observations, i.e., $p_{x(2t+1)|\bm{z}(2t), \mathcal H_0}$ the \ac{SAR} has been computed in \cite{6204019}. In particular, let us define $\bm{S} = \bm{R}_{[x(1) \bm{z}(2t+1)]}^{-1}$ and $\bm{T} =  \bm{R}_{[x(2t+1) \bm{z}(2t+1)]}^{-1}$ where
\begin{equation}
\begin{split}
\bm{R}&_{[x(2t+1) \bm{z}(2t+1)]} =\mathbb E[[x(2t+1), \bm{z}(2t)]^H[x(2t+1), \bm{z}(2t)]] \\
= &  \left[\begin{matrix}
1 +\sigma_{\rm x}^2 & \bm{r}_{\rm x,2}^H \\
\bm{r}_{\rm x,2} & \bm{R}_{\bm{z}(2t)}
\end{matrix}\right]\,,
\end{split}
\end{equation}
\begin{equation}
[\bm{r}_{\rm x,2}]_\ell = [\mathbb{E}[x(2t+1)\bm{z}^H(2t)]]_\ell = \begin{cases}
\alpha_{\rm A}^*\rho(\ell - 2t+1) & \ell \mbox{ even}\,, \\
\alpha_{\rm B}^*\rho(\ell - 2t+1) & \ell \mbox{ odd}\,. \\
\end{cases}
\end{equation}
Partitioning the two matrices as 
\begin{equation}
\bm{S} = \left[\begin{matrix}
S_{1,1} & \bm{S}_{1,2} \\
\bm{S}_{2,1} & \bm{S}_{2,2} \end{matrix}\right] \quad
\bm{T} = \left[\begin{matrix}
T_{1,1} & \bm{T}_{1,2} \\
\bm{T}_{2,1} & \bm{T}_{2,2} \end{matrix}\right]
\end{equation}
where $T_{1,1}$ and $S_{1,1}$ are scalars, while all other entries are vectors and matrices of suitable dimensions, we define 
\begin{equation}
\bm{E} = \bm{S}_{2,2} + \bm{T}_{2,2} - \bm{R}_{\bm{z}(2t+1)}^{-1}\,, 
\end{equation}
and 
\begin{equation}
\bm{V} = \left[\begin{matrix}
T_{1,1} - \bm{T}_{1,2}^H\bm{E}^{-1}\bm{T}_{1,2} &   -\bm{T}_{1,2}^H\bm{E}^{-1}\bm{S}_{1,2} \\
-\bm{S}_{1,2}^H\bm{E}^{-1}\bm{T}_{1,2}   & S_{1,1} - \bm{S}_{1,2}^H\bm{E}^{-1}\bm{S}_{1,2}
\end{matrix}\right]^{-1}\,.
\end{equation}
Then (\ref{rPLA1}) becomes
\begin{equation}
\begin{split}
R_{\rm PLA}& (2t+1) = \\
&\frac{-\ln \det(\bm{R}_{[x(2t+1), x(1)]}\bm{V}) + {\rm tr}(\bm{V}\bm{R}_{[x(2t+1), x(1)]}) - 2}{\ln 2}\,,
\end{split}
\end{equation}
where
\begin{equation}
\begin{split}
\bm{R}_{[x(2t+1), x(1)]} & = \mathbb E[[x(2t+1), x(1)]^H[x(2t+1), x(1)]] \\
& = \left[\begin{matrix}
1 + \sigma_{\rm x}^2 & \rho(2t) \\
\rho(-2t) & 1 + \sigma_{\rm x}^2
\end{matrix}\right]\,.
\end{split}
\end{equation}

\subsection{Linear Eve's Processing}
\label{suboptimal}

With \ac{LEP} Eve first performs a linear combination of all channel estimates to obtain the \ac{MMSE} linear estimate of the legitimate channel $\hat{h}$, which is then used for the attack. In particular, for the \ac{ACBCA} scheme Eve estimates $h(2)$, while for \ac{PLA} she estimates $h(1)$. For \ac{SCBCA} since we have only bounds on the \ac{SAR}, Eve can maximize the \ac{SAR} lower bound. Therefore, from the estimates obtained for \ac{PLA} and \ac{ACBCA} Eve picks the one that minimizes the minimum of the two mutual information, i.e.,
\begin{equation}
\hat{h} = {\rm argmin}_{h \in \{\hat{h}(1), \hat{h}(2)\}} \min \{{\mathbb I}(x(1);\bm{z}(2t)), {\mathbb I}(y(2);\bm{z}(2t))  \}\,.
\end{equation}

Let $k$ be the frame index of the desired channel estimate, $k= 1,2$. We first define the correlation vector
\begin{equation}
\bm{\beta}(k) = \mathbb{E}[\bm{z}(2t)h^*(k)]
\end{equation}
with entries
\begin{equation}
\beta_{2\ell+1}(k) = \alpha_{\rm A} \rho(k-2\ell-1)\,,  \quad \beta_{2\ell}(k) = \alpha_{\rm B} \rho(k-2\ell)\,. 
\end{equation}
Then we have 
\begin{equation}
\bm{z}(2t) = \bm{\beta}(k) h(k) + \bm{w}_0 \bm{R}_{\rm s}^{1/2}(2t) \,,
\end{equation}
where $\bm{w}_0$ is a jointly \ac{ZMUPCG} $2t$-size vector with \ac{iid} entries and the correlation matrix $\bm{R}_{\rm s}(2t)$ is
\begin{equation}
\bm{R}_{\rm s}(2t) = \mathbb{E}[(\bm{z}(2t) - \bm{\beta}(k) h(k))^H(\bm{z}(2t) - \bm{\beta}(k) h(k))]\,.
\end{equation}
Now, the \ac{MMSE} estimation of $h(k)$ is obtained as follows
\begin{equation}
\hat{z}(2t) = \frac{1}{{\rm tr}(\bm{R}_{\rm s}^{-2}(2t))} \bm{z}(2t)\bm{R}_{\rm s}^{-1}(2t)\bm{1} = h(k) + \sigma_{\rm z}(2t) \hat{w}_{\rm z}(2t)\,,
\end{equation}
where $\bm{1}$ is a $2t$-long column vector of all ones, $\hat{w}_{\rm z}(2t)$ is \ac{ZMUPCG} and 
\begin{equation}
\sigma_{\rm z}^2(2t) = {\rm tr}^2(\bm{R}_{\rm s}^{-2}(2t))\,.
\end{equation}

In general, \ac{LEP} is a suboptimal procedure (except for time-invariant channels) in the sense that the attacks by Eve will be less effective. However, this procedure has a limited computational complexity. The \ac{SAR} obtained with \ac{LEP} is readily computed from the results of the previous section where $\bm{z}(2t)$ is replaced by $\hat{z}(2t)$ having unitary correlation with $h(k)$ and noise power $\sigma_{\rm z}^2(2t)$.

\subsection{\ac{LEP} with Time-invariant Channels}
\label{LEPTI}

We now focus on time invariant channels, therefore $h(t) = h$, $q_{\rm A}(t)=q_{\rm A}$ and $q_{\rm B}(t) = q_{\rm B}$. We also have $\rho(t)=1$ $\forall t$. In this case, the \ac{LEP} procedure corresponds to optimal processing at Eve. 

For $t=1$ we have 
\begin{equation}
\bm{\beta}(1) = \bm{\beta}(2) = [\alpha_{\rm A}, \alpha_{\rm B}]^T
\end{equation}
and
\begin{equation}
\bm{R}_s = \left[ \begin{matrix}
1 - \alpha_{\rm A}^2 +\sigma_{\rm v,A}^2 & 0 \\
0 & 1 - \alpha_{\rm B}^2 + \sigma_{\rm v,B}^2 \\ 
\end{matrix}\right]\,.
\end{equation}
For the upper bound of \ac{SCBCA} \ac{SAR} we have 
\begin{equation}
\bm{R}_{[x(1), y(2), \hat{z}(2t)]} = \left[\begin{matrix}
1 + \sigma_{\rm x}^2 & 1 & 1 \\
1 & 1 + \sigma_{\rm y}^2 & 1 \\
1 &  1 &   1 + \sigma_{\rm z}^2(2t) \end{matrix}\right]\,,
\end{equation}
\begin{equation}
\begin{split}
\det\, & \bm{R}_{[x(1), y(2), \hat{h}(2)]} = (1 + \sigma_{\rm x}^2)[(1 + \sigma_{\rm y}^2)(1 + \sigma_{\rm z}^2(2)) -1] - \\
 &[(1 + \sigma_{\rm z}^2(2))-1] + [1-(1 + \sigma_{\rm y}^2)] \\
 & =  \sigma_{\rm x}^2\sigma_{\rm y}^2+  \sigma_{\rm x}^2\sigma_{\rm z}^2(2) + (1+\sigma_{\rm x}^2)\sigma_{\rm y}^2\sigma_{\rm z}^2(2).
\end{split}
\end{equation}

For $t>1$ note instead that, since $q_{\rm A}$ and $q_{\rm B}$ are the same at all frames, \ac{LEP} boils down to first estimating
\begin{equation}
\begin{split}
\bar{v}_{\rm A}(2t) = \frac{1}{t} \sum_{n=0}^{t-1} v_{\rm A}(2n+1) = \\
\alpha_{\rm A} h + \sqrt{1-\alpha_{\rm A}^2} q_{\rm A} + \sigma_{\bar{v},\rm A}(2t) w_{\rm v,A}(2t)\,,
\end{split}
\end{equation}
\begin{equation}
\begin{split}
\bar{v}_{\rm B}(2t) = \frac{1}{t} \sum_{n=0}^{t-1} v_{\rm B}(2n) = \\
\alpha_{\rm B} h + \sqrt{1-\alpha_{\rm B}^2} q_{\rm B} + \sigma_{\bar{v},\rm B}(2t) w_{\rm v,B}(2t)\,,
\end{split}
\end{equation}
where $w_{\rm v,A}(2t)$ and $w_{\rm v,B}(2t)$ are \ac{ZMUPCG} and 
\begin{equation}
\sigma_{\bar{v},\rm A}^2(2t) = \frac{\sigma_{\rm v,A}^2}{t}\,, \quad \sigma_{\bar{v},\rm B}^2(2t) = \frac{\sigma_{\rm v,B}^2}{t}\,,
\label{svt}
\end{equation}
and then applying \ac{MMSE} combining on $\bar{v}_{\rm A}(2t)$ and $\bar{v}_{\rm B}(2t)$ as for the case $t=1$.

In particular, for $t\rightarrow \infty$, from (\ref{svt}) we have that $\sigma_{\bar{v},\rm A}^2(2t) = \sigma_{\bar{v},\rm B}^2(2t) = 0$, and 
\begin{equation}
\hat{z}(2t) =  h + \frac{\alpha_{\rm A}(1-\alpha_{\rm A})}{(\alpha_{\rm A}^2 + \alpha_{\rm B}^2)} q_{\rm A} + \frac{\alpha_{\rm B}(1-\alpha_{\rm B})}{(\alpha_{\rm A}^2 + \alpha_{\rm B}^2)} q_{\rm B}\,,
\end{equation}
i.e., the Eve's channel estimate is affected only by $q_{\rm A}$ and $q_{\rm B}$.

In Appendix \ref{ap1} we derive the \ac{SAR} with \ac{LEP} processing of the  \ac{ACBCA} scheme  considering a uniform quantizer with saturation interval $[-v_{\rm sat}, v_{\rm sat}]$ and quantization step $\Delta$.

\section{Attacks  in the \ac{IDA} Phase}
\label{attackana}

In this section we consider attacks by Eve in the \ac{IDA} phase for the various \ac{UA} strategies. Eve transmits together and synchronously with Alice and Bob in the \ac{IDA} phase. Therefore she can overlap her signal on the pilots transmitted by Alice. Furthermore, Eve is assumed to be a full-duplex terminal, therefore she can transmit pilots and at the same time receive signals by Alice and Bob thus estimating the channel. Channels are time-invariant (thus we drop index $t$) and, in order to simplify notation, we assume that Eve has perfect estimates of her channels to both Alice and Bob, i.e., $\sigma_{\rm v}^2 =0$. 

Two attacks are considered: \ac{PC} and \ac{AN} attack. These attacks are very well known in the literature, and we now apply them to the \ac{UA} procedures. 
 
\paragraph{\ac{PC} Attack} With \ac{PC} attack, Eve transmits a scaled version of pilots transmitted by Alice, with scaling factors $\zeta_{\rm A}$ and $\zeta_{\rm B}$, so that Alice and Bob estimate the same channel to Eve, i.e.,
\begin{equation}
\zeta_{\rm A} v_{\rm A} = \zeta_{\rm B}v_{\rm B} = G\,.
\end{equation}
The estimated channels by Alice and Bob are
\begin{equation}
x(t) = h + G + \sigma_{\rm x} w_{\rm x}(t),\; y(t) = h + G + \sigma_{\rm y} w_{\rm y}(t)\,.
\end{equation}
In order to analyze the performance of this attack we note that if we divide $x(t)$ and $y(t)$ by $\sqrt{1 + |G|^2}$ we obtain again the model (\ref{xeq})-(\ref{veqB}) where now $\sigma_{\rm x}^2$ and $\sigma_{\rm y}^2$ become $\sigma_{\rm x}^2/(1 + |G|^2)$ and $\sigma_{\rm y}^2/(1 + |G|^2)$, respectively. On her side, Eve using \ac{LEP} obtains the estimate of $(h+G)/\sqrt{1 + |G|^2}$ 
\begin{equation}
\hat{z}_{\rm PC}(2t) =  (\hat{z}(2t) + G)/\sqrt{1 + |G|^2}
\end{equation}
with noise variance $\sigma_{\rm z}^2(2t)/(1 + |G|^2)$. Therefore the effect of this attack is the scaling of all noise variances, and results of previous Section can be used to compute the \ac{SAR}. Note that the \ac{PC} attack has an impact also on \ac{PLA} correctness, when Eve does not transmit pilots after the \ac{IDA} phase. In this case Bob may not recognize the Alice-Bob channel as correct, since $G$ is missing. \ac{ACBCA} and \ac{SCBCA} are not affected by this issue, since they only use the key extracted in the \ac{IDA} phase.

\paragraph{\ac{AN} Attack} With this attack Eve transmits \ac{AN} during the \ac{IDA} phase, i.e. a random \ac{ZMUPCG} signal aimed at increasing the noise for Alice and Bob. This scenario can be analyzed using the results of the previous section, simply modifying the values of $\sigma_{\rm x}^2$ and $\sigma_{\rm y}^2$. Note that this attack has no impact on the correctness of \ac{UA} process.

\subsection{Defense Strategies} 

We describe now possible defense strategies against the \ac{IDA} attacks.
\begin{itemize}
\item {\bf Random pilots}: legitimate parties use random pilots, locally generated at the transmitter and shared with the legitimate receiver after the \ac{IDA} phase on the public authenticated error-free channel (which as we have seen, must be in any case available in the \ac{IDA} phase). In this case Eve would not be able to add coherently her pilot and induce a desired channel;
\item {\bf Channel and noise power estimation}: if reference values of these powers are available at the legitimate receivers, the attack can be detected (see also \cite{Tomasin-glc16});
\item {\bf Channel agreement}: as outlined in \cite{Tomasin-glc16} by estimating the channel at both Alice and Bob and comparing the estimates without disclosing them to Eve it is possible to check if the two legitimate users see the same channel, thus preventing Eve from performing an attack in which she does not know the channels to Alice and Bob.
\end{itemize}
Moreover, note that the described attacks require the knowledge of the Alice-Eve and Bob-Eve channels before transmissions, therefore implementing the \ac{IDA} stage at the very beginning of transmission would prevent Eve from getting the channel estimates and deploy the attack.

\section{Numerical Results}

We provide now some results on the \ac{SAR} of the various \ac{UA} systems. We focus in particular on the Rayleigh \ac{AWGN} reciprocal channels of Section \ref{secAWGN}, where both Alice and Bob transmit with unitary power and channels are vectors of \ac{iid} \ac{ZMUPCG}. We consider both time-invariant and time-variant channels, and \ac{IDA} attacks described in Section \ref{attackana}. For \ac{ACBCA} we have already observed that the \ac{SAR} can be made arbitrarily large in the presence of a passive eavesdropper: here we report the results for a uniform quantizer with 3 bits (corresponding to 8 quantization levels) and saturation value $v_{\rm sat}$ that ensures a probability of saturation of $10^{-2}$. Unless differently specified we consider $\sigma^2_{\rm x} = \sigma^2_{\rm y} = \sigma^2_{\rm v, A/B} = -10$~dB.

\subsection{Time-invariant Channels} 

\begin{figure}
\includegraphics[width=1\hsize]{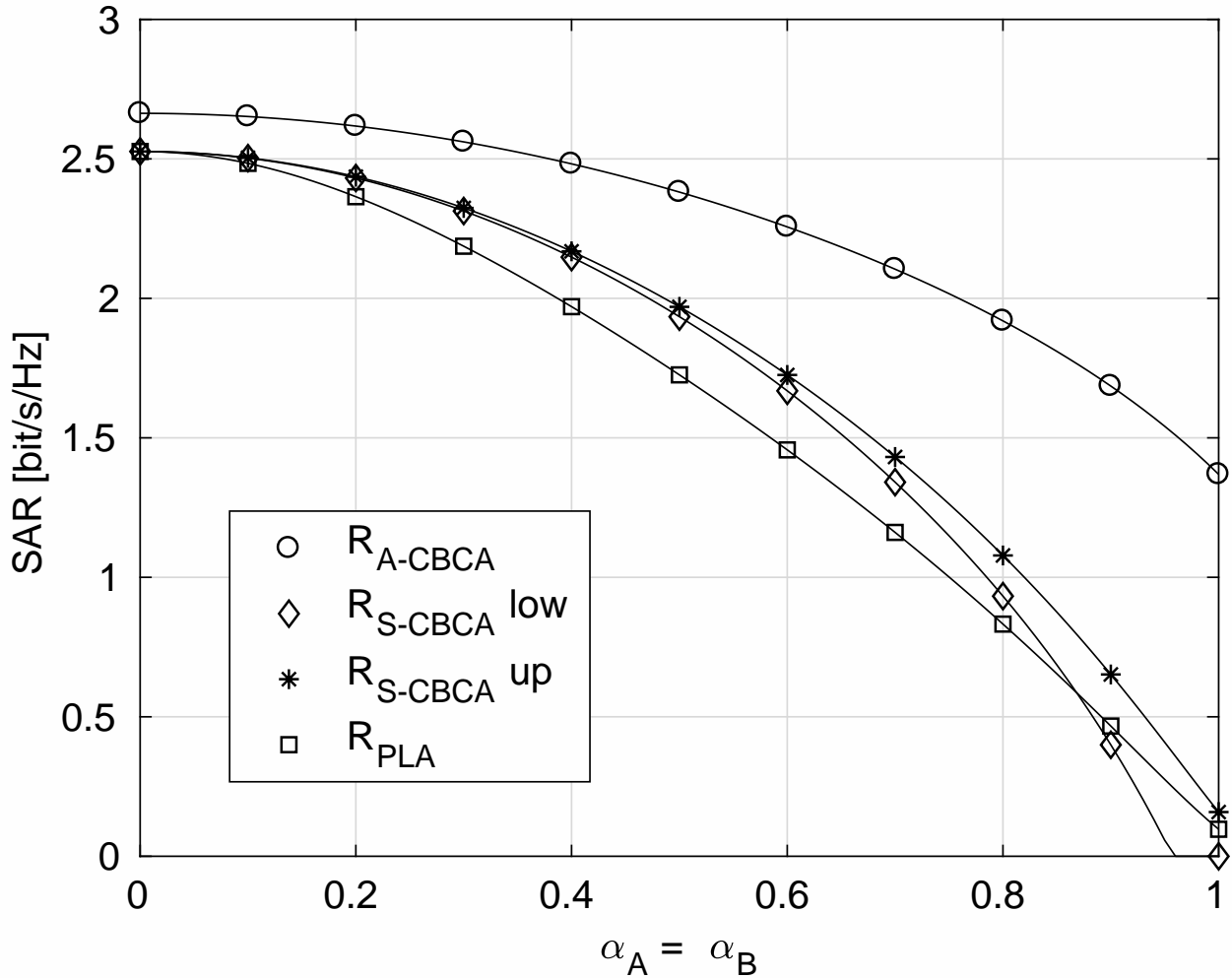}
\caption{\ac{SAR} vs the correlation coefficients $\alpha_{\rm A} = \alpha_{\rm B}$ for time-invariant channels at \ac{IDV} frame $t=3$.}
\label{fig1}
\end{figure}

We start from time-invariant channels. Fig. \ref{fig1} shows the \ac{SAR} versus (vs) the correlation coefficients $\alpha_{\rm A} = \alpha_{\rm B}$, at \ac{IDV} frame $t=3$, i.e., immediately after the two \ac{IDA} frames. The results for the lower and upper bound on the \ac{SAR} of \ac{SCBCA} are shown as $R_{\rm S-CBCA}$ low and up, respectively. As expected, with an increasing correlation among the legitimate and eavesdropper's channels the \ac{SAR} decreases. Note that even when $\alpha_{\rm A} = \alpha_{\rm B}=1$ we still may have a non-zero \ac{SAR}. In particular, in \ac{ACBCA} Bob benefits from the randomness of the noise which is assumed independent with respect to that of Eve. Similarly, in \ac{PLA} Eve generates a random attack channel having as mean her channel estimate rather than the estimate obtained by the legitimate user in the \ac{IDA} steps, and the two estimates differ due to the noise. The lower bound for \ac{SCBCA} is indeed zero in this case.

\begin{figure}
\includegraphics[width=1\hsize]{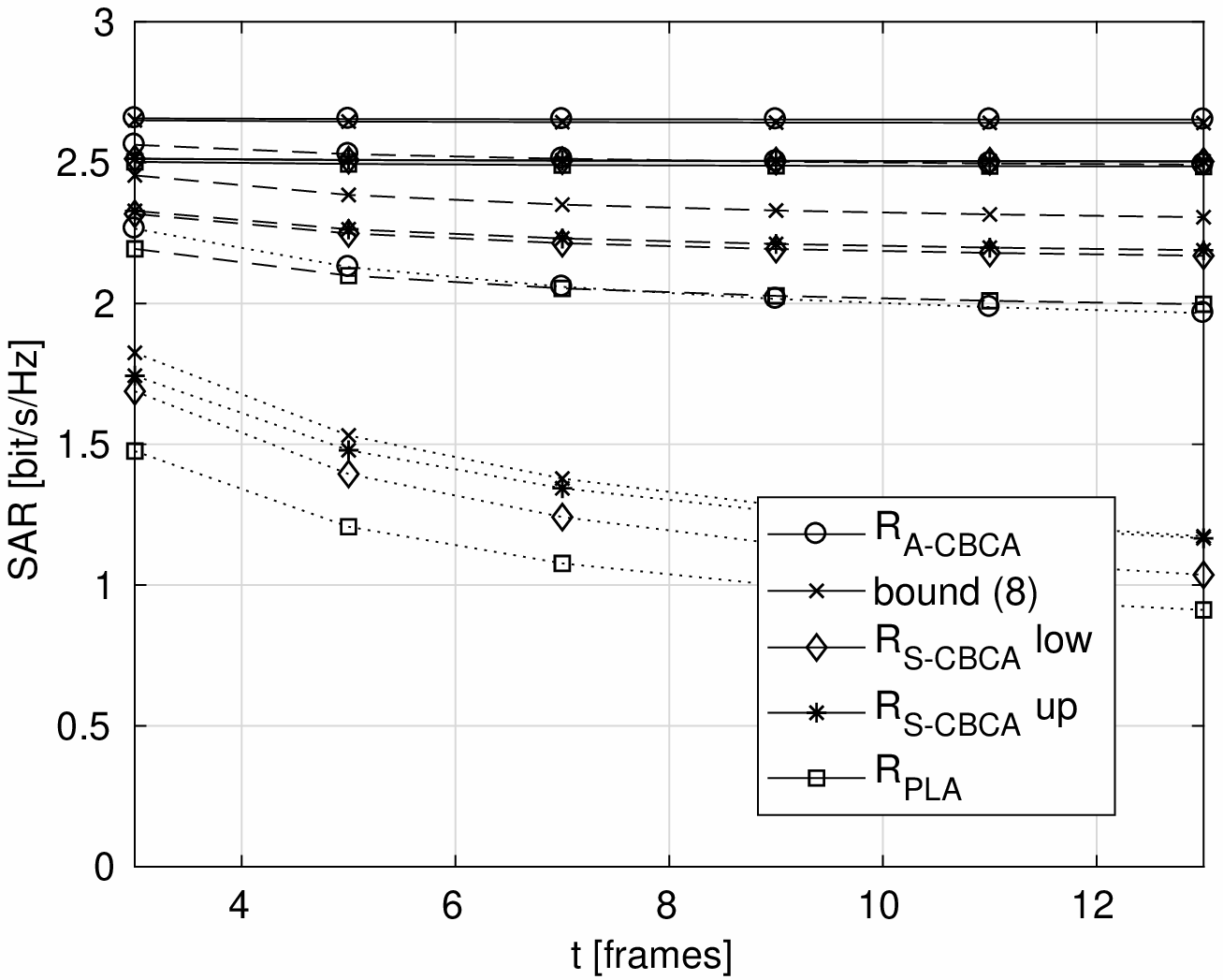}
\caption{\ac{SAR} vs \ac{IDV} frame index, for three values of channel correlations: $\alpha_{\rm A} = \alpha_{\rm B} = 0.1$ (solid lines), $\alpha_{\rm A} = \alpha_{\rm B} = 0.4$ (dashed  lines), and $\alpha_{\rm A} = \alpha_{\rm B} = 0.8$ (dotted lines).}
\label{fig3}
\end{figure}

Fig. \ref{fig3} shows the \ac{SAR} vs the \ac{IDV} frame for three values of channel correlations: $\alpha_{\rm A} = \alpha_{\rm B} = 0.1$ (solid lines), $\alpha_{\rm A} = \alpha_{\rm B} = 0.4$ (dashed  lines), and $\alpha_{\rm A} = \alpha_{\rm B} = 0.8$ (dotted lines). We observe that for all schemes, as the \ac{IDV} frame $t$ increases \ac{SAR} decreases, because in the meantime Eve has obtained a better channel estimate. The degradation of authentication performance is more remarkable for a higher value of channel correlation factor, since in this case having a more accurate knowledge of her channels to Alice and Bob truly provides Eve a better knowledge of the Alice-Bob channel.

\subsection{Time-varying Channels}

We consider now frame-time-variant channels with Jakes fading. In particular the channel is time-invariant in each frame while the evolution over frames is 
\begin{equation}
h(t) = \rho(t)h(1) + \sqrt{1 - |\rho(t)|^2} g(t)\,,
\end{equation}
with $g(t)$ \ac{ZMUPCG} and 
\begin{equation}
\rho(t) =  J_0(2\pi f_d tT)\,,
\label{bessel}
\end{equation}
with $T$ the frame duration, $f_d$ the Doppler frequency and $J_0(\cdot)$ the zero-order Bessel function of the first kind. 

\begin{figure}
\includegraphics[width=1\hsize]{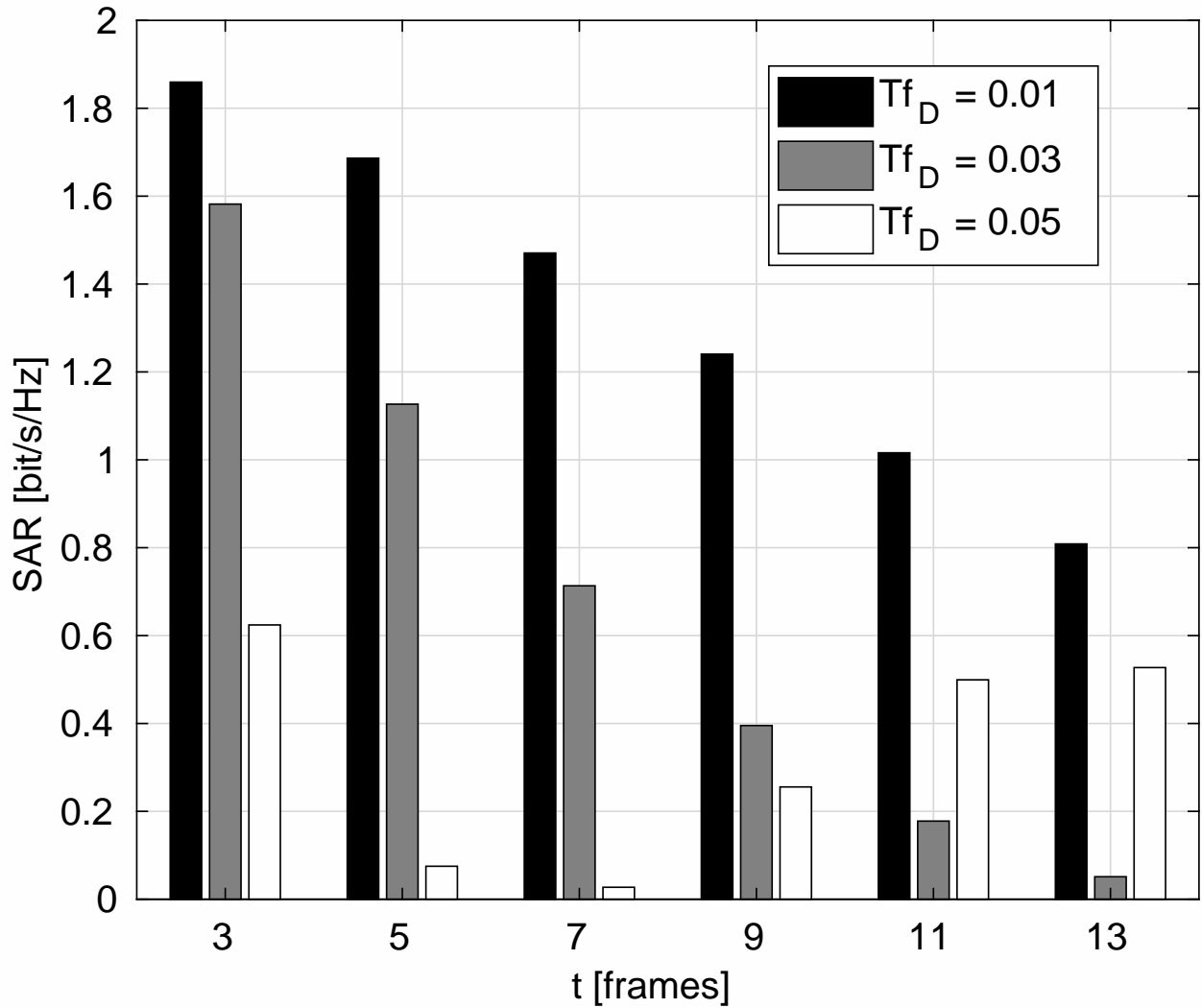}
\caption{\ac{SAR} vs the \ac{IDV} frame index, for three values of the normalized Doppler frequency $Tf_{\rm D}$ for the \ac{PLA} scheme. $\alpha_{\rm A} = \alpha_{\rm B} = 0.4$.}
\label{fig2}
\end{figure}
 
As we have observed, for \ac{PLA} channel variations have an impact on the ability of the scheme to effectively authenticate a legitimate transmission, since the channel (which is used as user signature) changes. Therefore Fig. \ref{fig2} shows the \ac{SAR} vs the \ac{IDV} frame index, for three values of the normalized Doppler frequency $Tf_{\rm D}$ for the \ac{PLA} scheme when $\alpha_{\rm A} = \alpha_{\rm B} = 0.4$. We assume that Eve  estimates the channel in only the first two frames, and up to frame $t>2$ Alice and Bob are not transmitting. We observe that as the normalized Doppler frequency increases the \ac{SAR} decreases. Moreover, as \ac{IDV} frame index increases the \ac{SAR} is reduced as well. In both cases the channel variations prevent an effective authentication. Lastly, note that for the highest value of the normalized Doppler frequency the \ac{SAR} increases for a higher number of frames: this is due to the Jakes model, and in particular to the behavior of the correlation (\ref{bessel}) as $f_d$ first decreases and then increases.

\begin{figure}
\includegraphics[width=1\hsize]{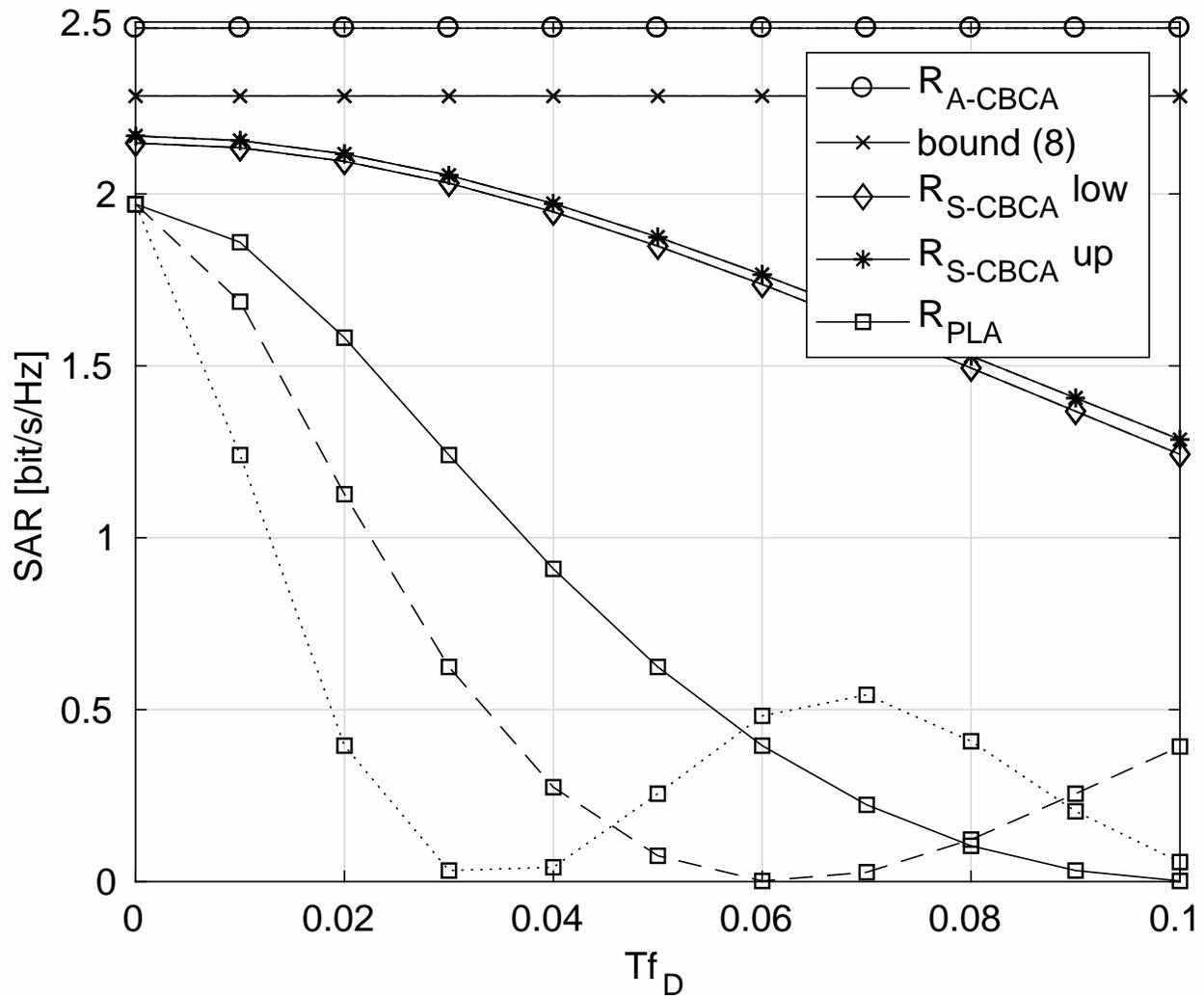}
\caption{\ac{SAR} vs the normalized Doppler frequency for three \ac{IDV} frames $t=3$ (solid lines), $t=5$ (dashed lines) and $t=9$ (dotted lines). $\alpha_{\rm A} = \alpha_{\rm B} = 0.4$.}
\label{fig4}
\end{figure}

We now consider the effect of time-variations on all the schemes. Fig. \ref{fig4} shows the \ac{SAR} as a function of the normalized Doppler frequency for three \ac{IDV} frames $t=3$ (solid lines), $t=5$ (dashed lines) and $t=9$ (dotted lines). Also in this case  $\alpha_{\rm A} = \alpha_{\rm B} = 0.4$. We observe that the \ac{ACBCA} scheme is not affected by the time-variation of the channel across frames, as it only uses one frame. Moreover, for \ac{SCBCA} the \ac{SAR} decreases for increasing normalized Doppler frequency but it is insensitive to the frame in which authentication is performed: in fact, for this scheme the channel is used only in the \ac{IDA} phase to establish the secret key and channel variations in further frames are not relevant. On the other hand, \ac{SCBCA} and \ac{PLA} are more heavily affected, since channel variations have an impact on both \ac{IDA} and \ac{IDV} phases. Also in this case we observe the effect of increasing channel correlation for high Doppler frequency, as already observed for Fig. \ref{fig3}.

\subsection{Active Attacks}

\begin{figure}
\includegraphics[width=1\hsize]{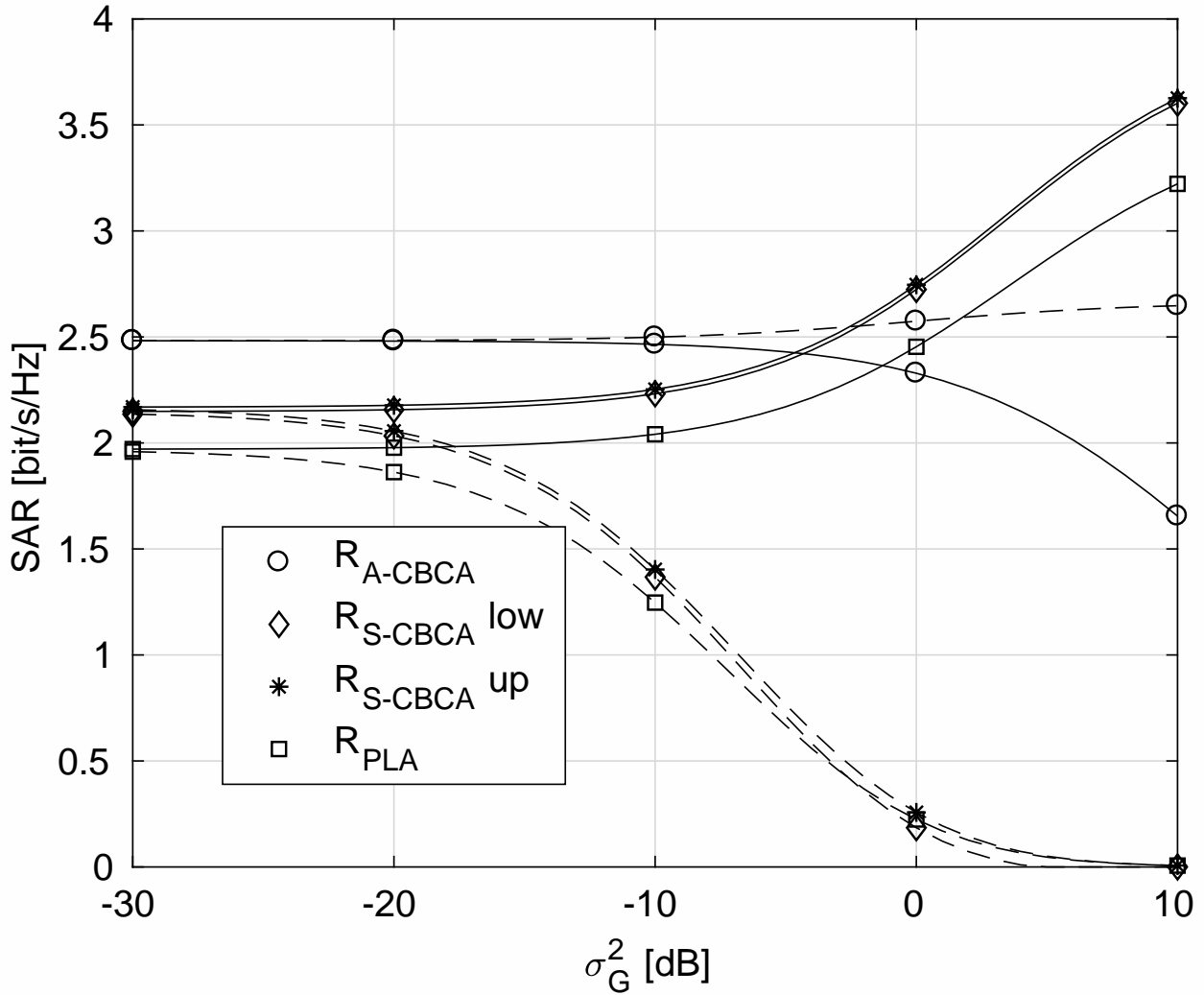}
\caption{\ac{SAR} as a function of $\sigma^2_G$ for \ac{IDA}-phase attacks, for both the \ac{PC} attack (solid lines) and the \ac{AN} attack (dashed lines).}
\label{fig5}
\end{figure}

We now consider the active attacks by Eve, as described in Section \ref{attackana}, namely \ac{PC} and \ac{AN} attack. Fig. \ref{fig5} shows the \ac{SAR} as a function of $\sigma^2_G$ for \ac{PC} attack (solid lines) and \ac{AN} attack (dashed lines). Also in this case $\alpha_{\rm A} = \alpha_{\rm B} = 0.4$ and channels are time-invariant. We recall that for the \ac{PC} attack $\sigma_{\rm G}^2$ is the power of the random channel superimposed to the effective channel that is estimated by the legitimate users in the \ac{IDA} phase. From the figure we observe that this attack is not effective for \ac{SCBCA} and \ac{PLA} schemes, instead increasing the \ac{SAR}. This is due to the fact that this attack is equivalent to a reduction of the noise estimate for both legitimate users and Eve, thus resulting in a globally favorable situation. For \ac{ACBCA} this attack in instead effective, since the reduction of the uncertainty on the channel estimated by Bob helps Eve to perform a more effective attack. When we consider the \ac{AN} attack instead we observe that it is effective for the \ac{SCBCA} and \ac{PLA} schemes, while it yields a higher \ac{SAR} (thus not being effective) for \ac{ACBCA}. In fact, since the \ac{AN} is not used by Eve in its estimation phase, this attack provides additional randomness to Alice, that is unknown to Eve, thus supporting the extraction of random bits from the channel in \ac{ACBCA}. Instead, for the other schemes the additional noise reduces the capabilities of extracting a shared secret key from the channel in the \ac{SCBCA} scheme, or the possibility of effectively detecting channel variations in \ac{PLA} scheme. Therefore this attack is effective against these two schemes. As we mentioned, randomizing pilots and channel agreement can be effective in preventing these attacks.

\section{Conclusions}

In this paper we have compared three \ac{UA} strategies, based on either symmetric/asymmetric keys or on physical layer authentication, where in all cases the features used for authenticating the user are extracted from the communication channel. The comparison has been performed in terms of \ac{SAR}, i.e., the rate at which the probability that \ac{UA} security is broken goes to zero as the number of used channel resources goes to infinity. From the analysis and the numerical results we can conclude that the \ac{ACBCA} scheme provides potentially the highest \ac{SAR} and is immune to channel changes. Then, the \ac{SCBCA} scheme, which uses the source-based \ac{SKA}, is slightly more sensitive to the channel variations but has a lower bound on \ac{SAR} that typically is higher than the \ac{SAR} achieved with \ac{PLA}. Moreover, \ac{PLA} is the most sensitive solution to channel variations.

\bibliographystyle{IEEEtran}

\bibliography{references}

\appendices

\section{Proof of Lemma \ref{lpr}}
\label{proof2}

We first observe that by the chain rule
\begin{equation}
\begin{split}
{\mathbb D}(p_{x(1),x(2t+1)|\mathcal H_0}||p_{x(1),x(2t+1)|\mathcal H_1}) \leq  \\
{\mathbb D}(p_{x(1),x(2t+1), \bm{z}(2t)|\mathcal H_0}||p_{x(1),x(2t+1), \bm{z}(2t)|\mathcal H_1})\,.
\end{split}
\label{Dpr}
\end{equation}
Moreover, the \ac{KL} divergence of the joint \acp{PDF} can be written as the expectation of the conditioned \acp{PDF}, i.e.,
\begin{equation}
\begin{split}
{\mathbb D}(p_{x(1),x(2t+1), \bm{z}(2t)|\mathcal H_0}||p_{x(1),x(2t+1), \bm{z}(2t)|\mathcal H_1}) = \\
{\mathbb E}[{\mathbb D}(p_{x(1),x(2t+1)|\mathcal H_0, \bm{z}(2t)}||p_{x(1),x(2t+1)|\mathcal H_1, \bm{z}(2t)})]\,,
\end{split}
\end{equation}
where expectation is taken with respect to $\bm{z}(2t)$. Now, we also have the general relation between mutual information and \ac{KL} divergence for  random variables $a$, $b$ and $c$
\begin{equation}
{\mathbb I}(a;b|c) =  
 {\mathbb E}_{c}[{\mathbb D}(p_{a,b|c}||p_{a|c}p_{b|c})].
 \label{Dpr2}
\end{equation} 
Recalling that conditionally on $\bm{z}(2t)$, $x(1)$ and $x(2t+1)$ (as generated by Eve) are independent \cite{7010914} we have
\begin{equation}
p_{x(1),x(2t+1)|\mathcal H_1, \bm{z}(t)} = p_{x(1)|\mathcal H_1, \bm{z}(t)}p_{x(2t+1)|\mathcal H_1, \bm{z}(2t)}\,.
\end{equation}
Now, assuming that Eve does not attack in the \ac{IDA} frames, $p_{x(1)|\mathcal H_1, \bm{z}(t)} = p_{x(1)|\mathcal H_0, \bm{z}(t)}$. Moreover, if the attack has \ac{PDF} $p_{x(2t+1)|\mathcal H_0,\bm{z}(2t)}$, we have 
\begin{equation}
p_{x(2t+1)|\mathcal H_1, \bm{z}(2t)} = p_{x(2t+1)|\mathcal H_0, \bm{z}(2t)}
\end{equation}
and therefore 
\begin{equation}
p_{x(1),x(2t+1)|\mathcal H_1, \bm{z}(t)} = p_{x(1)|\mathcal H_0, \bm{z}(t)}p_{x(2t+1)|\mathcal H_0, \bm{z}(2t)}\,,
\end{equation}
and using (\ref{Dpr})  we have 
\begin{equation}
\begin{split}
{\mathbb D}(p_{x(1),x(2t+1)|\mathcal H_0}||p_{x(1),x(2t+1)|\mathcal H_1})  \leq  \\
{\mathbb I}(x(1);x_{\mathcal H_0}(2t+1)|\bm{z}(2t))\,,
\end{split}
\end{equation}
which provides (\ref{rPLA}).

\section{\ac{SAR} for \ac{ACBCA} with Rayleigh fading}
\label{ap1}

We first observe that the Rayleigh channel coefficient $h$ has \ac{iid} real and imaginary parts, therefore the \ac{SAR} will be twice of the rate obtained considering the quantization of the real part, i.e., 
\begin{equation}
\begin{split}
R_{\rm A-CBCA}(2t+1) & = -2\sum_{i=0}^{M-1} \int p_{\langle y(2)\rangle|\hat{z}(2t)}(a_i|b) p_{\hat{z}(2t)}(b) \\
& \log p_{\langle y(2)\rangle|\hat{z}(2t)}(a_i|b)  {\rm d}b\,,
\end{split}
\label{Hgauss}
\end{equation}

For time-invariant channels and considering $(T_{i-1}, T_i)$ as quantization interval for $a_i$ we have 
\begin{equation}
\begin{split}
p&_{\langle y(2)\rangle|\hat{z}(2t)}(i|b) = \frac{1}{p_{\hat{z}(2t)}(b)} \times \\
\mathbb{P}&\left(\hat{h}+ \frac{\sigma_{\rm y}}{\sqrt{2}} w_{\rm y}(2) \in (T_{i-1}, T_i], \hat{h}+ \frac{\sigma_{\rm z}(2t)}{\sqrt{2}} \hat{w}_{\rm z}(t) =b, h=\hat{h}\right)\\
= &  \int_{-\infty}^\infty \mathbb{P}\left(h+ \frac{\sigma_{\rm y}}{\sqrt{2}} w_{\rm y}(2) \in (T_{i-1}, T_i]\right)  \times \\
& p_{\hat{w}_{\rm z}(2t)}	\left(-\frac{h\sqrt{2}}{\sigma_{\rm z}(2t)}+ \frac{b\sqrt{2}}{\sigma_{\rm z}(2t)}\right) p_{h}(h) dh\,,
\end{split}
\label{condyz}
\end{equation}
where we used  half of all noise variances since we are considering only the real part of the channel estimates. Now considering $v_{\rm sat}$ as the saturation value and $T_i = v_{\rm sat} + \Delta i$, $T_{-1} = -\infty$, $T_{M} = \infty$, we have that the first function in (\ref{Hgauss}) is 
\begin{equation}
\begin{split}
\mathbb{P}&\left(h+ \frac{\sigma_{\rm y}}{\sqrt{2}} w_{\rm y}(2)\in (T_{i-1}, T_i]\right) = \\
& \begin{cases}
1 - {\rm Q}\left(\frac{\sqrt{2}(-v_{\rm sat}+ \Delta  - h)}{\sigma_{\rm{y}}}\right) & i = 0 \\
{\rm Q}\left(\frac{\sqrt{2}(-v_{\rm sat}+ \Delta i - h)}{\sigma_{\rm{y}}}\right)
- {\rm Q}\left(\frac{\sqrt{2}(-v_{\rm sat}+ \Delta (i+1	- h)}{\sigma_{\rm{y}}}\right) & i \in [1, M -2]\\
{\rm Q}\left(\frac{\sqrt{2}(-v_{\rm sat}+ \Delta (M-1)- h)}{\sigma_{\rm{y}}}\right) & i = M-1\,, \\
\end{cases} \\
\end{split}
\label{Qexp}
\end{equation}
where $\langle y(2)\rangle$ is the quantization index of $y(2)$. Moreover for second and third functions in (\ref{Hgauss}) we have
\begin{equation}
p_{h}(\hat{h}) = \frac{1}{ \pi}e^{-\hat{h}^2}\,, 
\end{equation}
\begin{equation}
p_{\hat{w}_{\rm z}(2t)}\left(-\frac{\sqrt{2}h}{\sigma_{\rm z}(2t)}+ \frac{\sqrt{2}b}{\sigma_{\rm z}(2t)}\right) = \frac{1}{\sigma_{\rm z}(2t)\sqrt{ \pi}}e^{-\frac{(h-b)^2}{ \sigma_{\rm z}^2(2t)}}.
\end{equation}
Both integrals in (\ref{condyz}) and in (\ref{Hgauss}) must be solved by numerical methods.
%

\end{document}